# Towards the growth of single-crystal boron nitride monolayer on Cu

Li Wang[1,2,#], Xiaozhi Xu[1,#], Leining Zhang[3,4,#], Ruixi Qiao[1,#], Muhong Wu[1,5], Zhichang Wang[6], Shuai Zhang[7], Jing Liang[1], Zhihong Zhang[1], Yuwei Shan[8], Yi Guo[1], Marc Willinger[9,10], Hui Wu[11], Qunyang Li[7], Wenlong Wang[2], Peng Gao[6], Shiwei Wu[8], Ying Jiang[6], Dapeng Yu[12,13], Enge Wang[6], Xuedong Bai[2]*, Zhu-Jun Wang[9,10]*, Feng Ding[3,4]*, Kaihui Liu[1]*

[1] State Key Laboratory for Mesoscopic Physics, Collaborative Innovation Center of Quantum Matter, School of Physics, Peking University, Beijing 100871, China

[2] Beijing National Laboratory for Condensed Matter Physics, Institute of Physics, Chinese Academy of Sciences, Beijing 100190, China

[3] Centre for Multidimensional Carbon Materials, Institute for Basic Science, Ulsan 44919, Korea

[4] School of Materials Science and Engineering, Ulsan National Institute of Science and Technology, Ulsan 44919, Korea

[5] Songshan Lake Laboratory for Materials Science, Dongguan 523808, China

[6] International Center for Quantum Materials, School of Physics, Peking University, Beijing 100871, China

[7] Department of Engineering Mechanics, State Key Laboratory of Tribology, Tsinghua University, Beijing 100084, China

[8] State Key Laboratory of Surface Physics, Department of Physics, Fudan University, Shanghai 200433, China

[9] Scientific Centre for Optical and Electron Microscopy, Eidgenössische Technische Hochschule Zürich, 8092 Zürich, Switzerland

[10] Department of Inorganic Chemistry, Fritz Haber Institute of Max Planck Society, Berlin 14195, Germany

[11] State Key Laboratory of New Ceramics, Fine Processing School of Materials Science and Engineering, Tsinghua University, Beijing 100084, China

[12] Institute for Quantum Science and Engineering and Department of Physics, South University of Science and Technology of China, Shenzhen 518055, China

[13] Shenzhen Key Laboratory of Quantum Science and Engineering, Shenzhen 518055, China

# These authors contributed equally to this work
* Correspondence: khliu@pku.edu.cn; f.ding@unist.ac.kr; zhujun@fhi-berlin.mpg.de; xdbai@iphy.ac.cn

**Atom-layered hexagonal boron nitride (hBN), with excellent stability, flat surface and large bandgap, has been reported to be the best 2D insulator to open up the great possibilities for exciting potential applications in electronics, optoelectronics and photovoltaics[1-17]. The ability to grow high-quality large single crystals of hBN is at the heart for those applications, but the size of single-crystal 2D BN is less than a millimetre till now[18-24]. Here, we report the first epitaxial growth of a 10 ×10 cm$^2$ single-crystal hBN monolayer on a low symmetry Cu(110) "vicinal surface". The growth kinetics, unidirectional alignment and seamless stitching of hBN domains are unambiguously illustrated using centimetre- to the atomic-scale characterization techniques. The findings in this work are expected to significantly boost the massive applications of 2D materials-based devices, and also pave the way for the epitaxial growth of broad non- centrosymmetric 2D materials[25-28].**

In the family of two-dimensional materials, 2D hexagonal boron nitride, a binary compound with a similar lattice structure to graphene, has attracted increasing interest due to its exceptional properties[7-17]. With regard to 2D materials-based device applications, a large choice of 2D conductors and semiconductors belonging to various materials families are available, while hBN is the exclusive choice for a 2D insulator, due to its high mechanical and thermal stabilities and ultra-flat surface. The large difference in potential between the B and N sublattices in hBN can be used to tune the properties of 2D materials on its surface, for example, second-generation Dirac cones, Hofstadter butterfly patterns and Mott insulators can be realised from graphene on hBN[10-13]. In addition, hBN is an optical material with a deep UV bandgap[14], enables single photon emission from point defects[15], and also has the high room-temperature proton conductivity[16]. However, to date, the superior properties of hBN have been demonstrated for exfoliated 2D hBN samples with typical domain sizes in the range of tens of microns, which is obviously too small for real applications. Future large-scale applications of hBN and 2D materials-based devices thus depend on the availability of wafer-size hBN, ideally in the form of single crystals, such as demonstrated for graphene[4-6].

Currently, the most prevalent method to produce 2D hBN is through chemical vapour deposition (CVD)[18-24] and to date, the maximum size of a single-crystal 2D hBN sample is less than 1 mm[22, 23]. This is about three orders of magnitude smaller than the largest graphene single crystal and is far smaller than that required in large scale applications. In principle, there are two ways to grow a 2D single crystal on a substrate: (i) By allowing a single nucleus to grow into a large single crystal on a substrate or (ii) by achieving unidirectional alignment of millions of nuclei on a substrate, guided by epitaxial growth. The first option is challenging in that it is nearly impossible to suppress multiple nucleation during hBN growth, and the second option requires a single-crystal substrate of an appropriate symmetry. It should be noted here that 2D single crystal growth through the second route can be orders of magnitude faster because of the simultaneous nucleation and growth of a large number of 2D islands.

Recently, we have demonstrated the synthesis of a macro-sized single-crystal Cu(111) foil and the epitaxial growth of single-crystal graphene on this substrate, which paved the way for the epitaxial growth of single crystals of various 2D materials (route ii). However, the Cu(111) surface is not an appropriate template for the growth of single-crystal hBN due to the nucleation of antiparallel hBN islands on this substrate[24]. This is because, the epitaxial growth of unidirectionally aligned islands requires that the 2D material precisely adopts the symmetry of the substrate to avoid change in lattice orientation during a symmetric operation of the substrate. Different from graphene, most 2D materials including hBN have a lower symmetry ($C_{3v}$) than graphene, and therefore are not compatible with the $C_{6v}$ symmetry of the top layer Cu atom of the Cu(111) surface. Thus, the ideal substrate for hBN growth must have the $C_{3v}$, $C_3$, $\sigma_V$ or $C_1$ symmetry. Since the Cu(111) surface cannot be used to grow hBN single crystal, we have to choose a substrate with only $\sigma_V$ or $C_1$ symmetry, which can definitely not be a regular low index face-centred cubic (fcc) surface. To address this challenge, we performed, in this study, the successful synthesis of a Cu(110) "vicinal surface", on which, the presence of metal steps along the <211> direction led to a $C_1$ symmetry. This allowed the unidirectional alignment of millions of hBN nuclei over a large 10 × 10 $cm^2$ area.

In our experiment, single-crystal foils of Cu(110) "vicinal surface" were prepared by directly annealing industrial Cu foils (Please see the section on Methods for more detail). When using our previously reported temperature-gradient-driven annealing technique[5], near-(110) domains in the raw polycrystalline Cu foil have a relatively high probability (~2% in current condition) to be evolutionarily expanded a large-size single crystal of 10 × 10 $cm^2$. The large-area Cu(110) surface can be easily observed by optical imaging after mild oxidization in air (Fig. 1a), as different surfaces have different oxidization rates to form $Cu_2O$ and show fingerprint colours[29]. The sharp Cu(220) peak in X-ray diffraction (XRD) pattern in the 2θ scan (Fig. 1b) and the presence of only two peaks at exactly 180° interval in the φ scan (Fig. 1c, fixed along Cu<100>) unambiguously confirmed that the single-crystal foil had no in-plane rotation. Furthermore, electron backscatter diffraction (EBSD) maps, low energy electron diffraction (LEED) patterns (Fig. 1e) and scanning transmission electron microscopy (STEM) images (Fig. 1f) at different positions revealed the single crystalline nature of the Cu(110) substrate. Here, we note that the exact surface index or the tilt angle from the <110>

direction can scarcely be observed, which means that the tilt angle is much smaller than the experimental accuracy of ~1°. We have chosen however, to call it Cu(110), unless otherwise specified.

Using the Cu foil single-crystal as the substrate, 2D hBN was synthesized by a low-pressure CVD method with ammonia borane ($H_3B$-$NH_3$) as the feedstock (Methods). X-ray photoelectron spectroscopy (XPS), Raman spectroscopy, UV-visible spectroscopy (UV-Vis) and atomic force microscopy (AFM) have been used to confirm that all the obtained domains were from a 2D hBN monolayer. SEM images showed a striking result, of hBN islands grown on the Cu(110) surface with all the triangular islands unidirectionally aligned on the Cu foil surface on the centimetre scale, and the ratio of the aligned islands was estimated to be ~ 99.5% (Fig. 2a). To confirm that the aligned islands had the same crystalline orientation, we performed LEED measurements. The LEED patterns (Fig. 2b) measured at randomly chosen positions confirmed that the crystalline lattice of hBN islands are aligned in the same direction [19]. The unidirectional alignment of hBN lattice was further proven by polarized second-harmonic generation (SHG) mapping[30], where a dark boundary line is observed between hBN lattices of different domains, when these are not aligned. As shown in Fig. 2c, no boundary line is observed in the bulk area of coalescence of hBN islands, also indicating similar crystalline orientation of the two coalesced grains.

During the growth of a 2D material on Cu surface, a seamless stitching is expected during the coalescence of two unidirectional aligned grains, since the perfect single crystalline lattice is always the most stable structure, as has been proven in the CVD growth of graphene [4, 5]. To further confirm seamless stitching of domains, hydrogen ($H_2$) etching was employed to visualize the different possible boundaries on the macro scale[5, 6]. Fig. 2d showed no "etching line" between unidirectionally aligned hBN domains; in contrast, for domains of different alignments, the etched boundary was clearly visualized (Fig. 2e). UV light oxidization was also carried out to expose possible grain boundaries, when present[4, 5]. Similar results were obtained, allowing to conclude that the large-area hBN film is a single crystal. High-resolution transmission electron microscopy (HRTEM) was used to characterize the quality of the stitching line between domains on the atomic scale. The as-grown hBN samples were transferred onto homemade single-crystal graphene TEM girds to observe the Moiré patterns in the hBN/Graphene heterostructure. It is well known that even a small difference in the rotation angle can lead to a big change in the Moiré pattern[31]. Consistent Moiré patterns collected at multiple places around the concave corner (shown in Fig. 2f-g) confirmed that the unidirectionally aligned hBN domains are seamlessly stitched into an intact piece of single crystal. On the other hand, a clear difference in the Moiré pattern and grain boundaries were observed in the area of confluence of two misaligned hBN domains.

We have used environmental SEM to study the growth dynamics *in situ,* with the aim gain an in-depth understanding of the kinetics of epitaxial growth of unidirectional hBN islands on the Cu(110). Our *in-situ* observations clearly proved that our Cu(110) single-crystal substrate is "vicinal" by existence of the steps, and that the step edges play a crucial role in the unidirectional alignment of hBN islands. SEM images (Fig. 3a-b) clearly showed that each hBN single crystal is nucleated near a step edge and one edge of the single crystal is tightly attached to the up-hill side of the step edge during the growth process and the single crystal propagated rapidly on the terrace between neighbouring step edges. Once one of its edges reached a neighbouring step edge in the down-hill direction, the propagation of the edge was arrested for a while[32, 33]. Our *in-situ* observation thus clearly confirmed that the unidirectional alignment of hBN islands is caused by step edge-mediated nucleation and the truncated shape of the hBN islands results from the barrier of a hBN edge to climb over a passing step edge, both in the up-hill and in the down-hill directions. Due to the presence of parallel step edges (equal surface tilt angle) on the single-crystal Cu(110) surface, this growth kinetics leads to unidirectionally aligned hBN domains (Fig. 3c); the process is schematically illustrated in Fig. 3d.

In conclusion, $10 \times 10$ cm$^2$ area single-crystal hBN films that are three orders of magnitude larger than observed up to now, were synthesized on a large-area single-crystal Cu(110) foil obtained by annealing regular industrially produced Cu foils. The easy preparation procedure for the Cu substrate implies the immediate availability and wide application of single-crystal 2D hBN films. This method is expected to be applicable to all non-centrosymmetric 2D materials in general, which will enable the growth of large area single crystals of these materials and facilitate their use in different applications in the near future.

**Figures and Captions:**

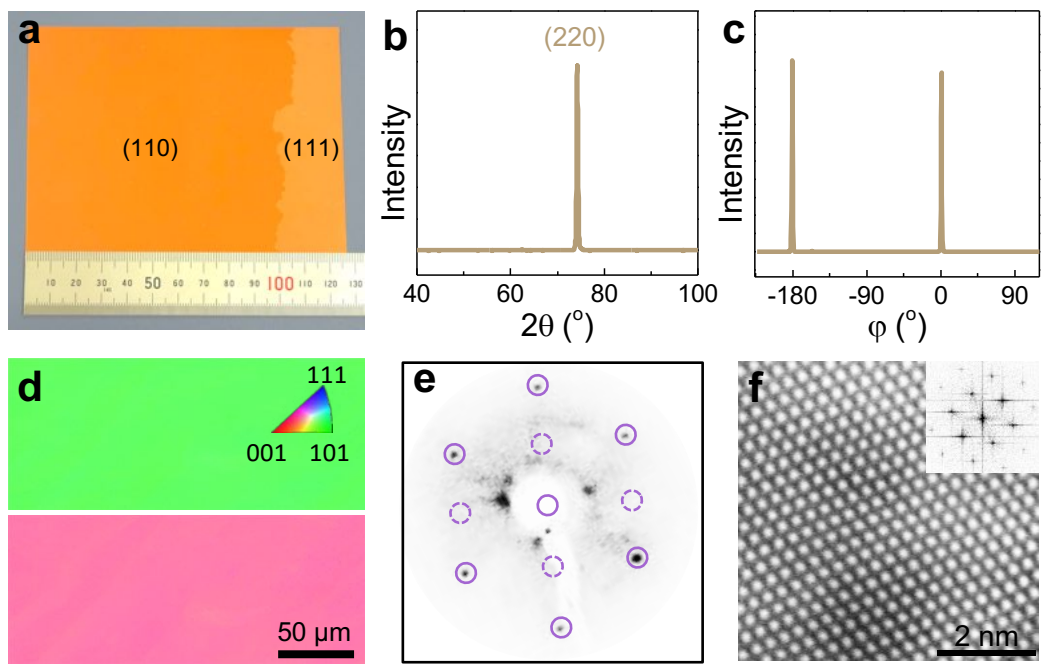

**Figure 1 | Characterization of single-crystal Cu(110) obtained by annealing an industrial Cu foil. a,** Optical image of the annealed Cu foil after mild oxidation in hot air. Due to the different oxidation rates, Cu(110) and Cu(111) domains have different $Cu_2O$ thickness on the top and therefore show different colours. With this method, typical $10 \times 10$ $cm^2$-sized Cu(110) single crystals were obtained. **b-c,** XRD pattern corresponding to 2θ scan of Cu(110) foil **(b)** and φ scan along the Cu(100) direction **(c),** confirming the single crystalline nature without in-plane rotation, of the Cu(110) foil. **d,** Representative EBSD maps of as-grown Cu(110) foils (upper panel: along [001] direction; lower panel: along [010] direction). **e,** Representative LEED pattern of as-grown Cu(110) foils. **f,** Atomically resolved STEM image of as-grown Cu(110) foils. Inset: Fast Fourier transformation pattern of the STEM image. Results from these characterization experiments together prove that the annealed Cu foil is single-crystal Cu(110).

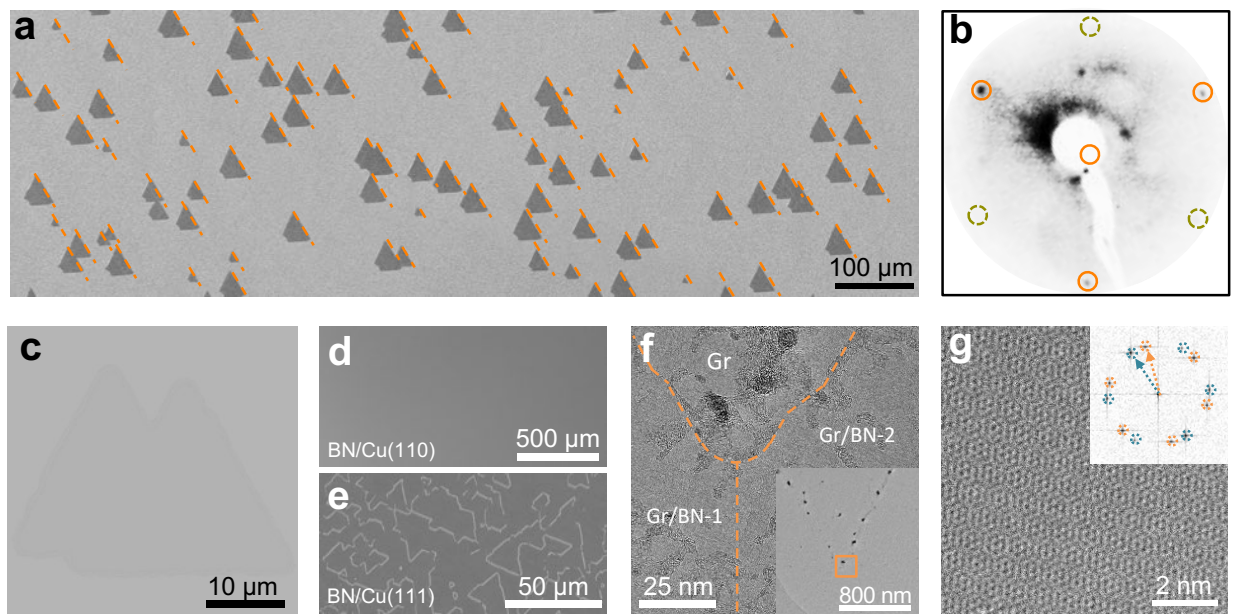

**Figure 2 | Unidirectional alignment and seamless stitching of hBN domains on Cu(110). a,** SEM image of as-grown unidirectionally aligned hBN domains on Cu(110) substrate. **b,** Representative LEED pattern of as-grown hBN samples. Due to the triple symmetry, three diffraction points have high intensity (marked by solid orange circles) and the other three are weaker in intensity (marked by dash green circles). **c,** Polarized SHG mapping of two unidirectionally aligned hBN domains. The uniform colour without boundary line demonstrates that the aligned hBN domains have the same lattice orientation. **d-e,** SEM images of as-grown hBN films after $H_2$ (250 sccm) etching at 1000 °C for 30 min. No boundaries are observed for hBN film grown on Cu(110) **(d)**, but obvious boundaries can be observed on Cu(111) **(e)**. **f,** Low magnification TEM image at the concave corner in the area of confluence of aligned hBN domains on monolayer single-crystal graphene support. The inset shows an image at a lower magnification. **g,** Representative HRTEM image showing the consistent Moiré pattern at the concave corner in the boundary area between unidirectionally aligned hBN domains. Insets: Fast Fourier transformation patterns.

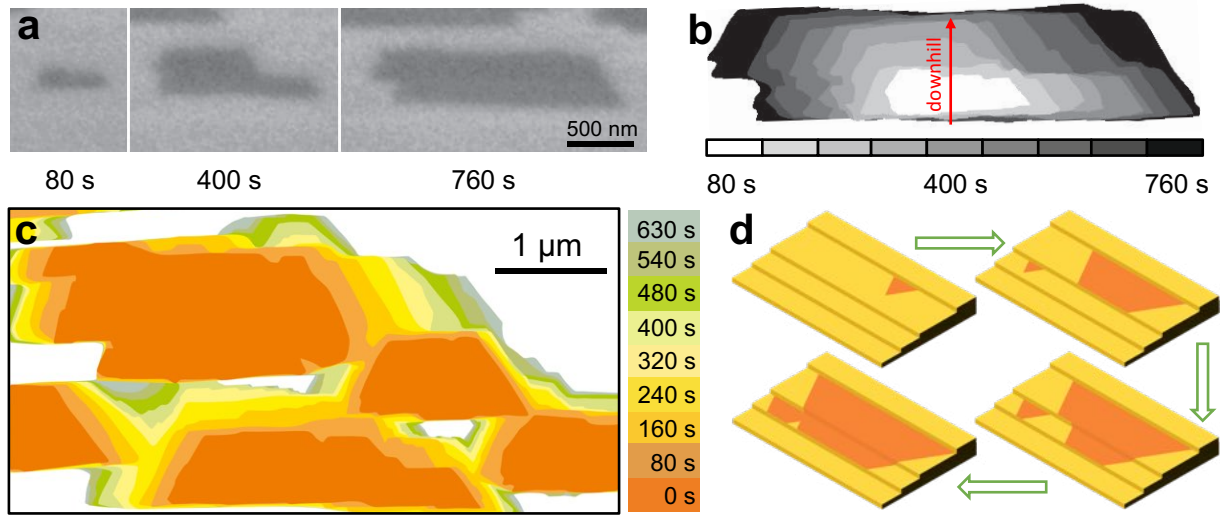

**Figure 3 | *In-situ* observation of the unidirectional growth of hBN domains. a,** *In situ* SEM images of a hBN domain at different growth times probably formed from a nucleation centre at the surface steps. **b,** Superposition capturing the areal growth of the domain in **(a)**. **c,** Shape evolution of several hBN domains, reproduced as a color-coded superposition of outlines extracted from images recorded during 700 s. **d,** Schematic diagrams highlighting the unidirectional growth of hBN domains and the anisotropic growth speed on a Cu surface with steps.